\documentclass[twocolumn]{aastex7}
\usepackage{color} 		
\usepackage{multirow}	
\usepackage{xspace}	    

\newcommand{\mvir}{$M_{\rm{vir}}$\xspace}
\newcommand{\rs}{$r_{\rm{s}}$\xspace}
\newcommand{\vlos}{$v_{\rm los}$\xspace}
\newcommand{\rproj}{$r_{\rm proj}$\xspace}
\newcommand{\cnnb}{$\rm{CNN}_{\rm{baseline}}$\xspace}
\newcommand{\cnnr}{$\rm{CNN}_{\rm{richness}}$\xspace}

\usepackage{physics}

\newcommand{\hMpc}{$ \, h^{-1}  \rm Mpc$\xspace}

\newcommand{\hMpccube}{$(h^{-1} \rm Mpc)^3$\xspace}

\newcommand{\hkpc}{$ \, h^{-1} \rm kpc$\xspace}

\newcommand{\hMsun}{$\, h^{-1} \rm M_\odot$\xspace}
\newcommand{\Msun}{$\, \rm M_\odot$\xspace}
\newcommand{\mtzzc}{$M_{\rm 200c}$\xspace}

\newcommand{\kms}{$\, \rm km \, s^{-1}$\xspace}
\newcommand{\lsim}{\mbox{${\,\hbox{\hbox{$ < $}\kern -0.8em \lower 1.0ex\hbox{$\sim$}}\,}$}}
\newcommand{\gsim}{\mbox{${\,\hbox{\hbox{$ > $}\kern -0.8em \lower 1.0ex\hbox{$\sim$}}\,}$}}
\newcommand{\SU}{{Skies \& Universes}\xspace}
\newcommand{\urluchuu}{\url{https://www.skiesanduniverses.org/Simulations/Uchuu/}}
\newcommand{\urlmrockstar}{\url{https://github.com/Tomoaki-Ishiyama/mpi-rockstar/}}

\definecolor{mygray}{gray}{0.7}

\begin{document}

\title{Inferring Halo Mass and Scale Radius of Galaxy Clusters Using Convolutional Neural Networks and Uchuu-UniverseMachine Catalogs}

\correspondingauthor{Tomoaki Ishiyama}

\author[0009-0007-4601-1513]{Hirobumi Tominaga}
\affiliation{Interdisciplinary Mathematical Sciences, Meiji University, 4-21-1, Nakano, Nakano-ku, Tokyo, 164-8525, Japan}
\email{}

\author[0009-0006-7479-0019]{Asuka Nakamura}
\affiliation{Chiba Institute of Technology, 2-17-1, Tsudanuma, Narashino, Chiba, 275-0016, Japan}
\email{}

\author[0000-0002-5316-9171]{Tomoaki Ishiyama}
\affiliation{Digital Transformation Enhancement Council, Chiba University, 1-33, Yayoi-cho, Inage-ku, Chiba, 263-8522, Japan}
\email[show]{ishiyama@chiba-u.jp}

\author[0000-0003-3595-7147]{Mohamed H. Abdullah}
\affiliation{Department of Physics, University of California Merced, 5200 North Lake Road, Merced, CA 95343, USA}
\affiliation{Department of Astronomy, National Research Institute of Astronomy and Geophysics, Cairo, 11421, Egypt}
\email{}

\begin{abstract}
We investigate the ability of machine learning to
infer the virial mass \mvir and the scale radius \rs of galaxy
clusters from their observables. Using the Uchuu--UniverseMachine
galaxy catalog at $z=0.093$, we generate mock cluster observations
that include interlopers, and we encode each cluster as an image
representing the two-dimensional joint probability distribution of
member galaxies' projected position and line-of-sight velocity. We
train two architectures: a baseline convolutional neural network (\cnnb) following a previous
approach, and an extended model (\cnnr) that appends richness as an
additional scalar input. We further compare the performance of networks
trained on the all cluster sample and on a dynamically relaxed
subsample.  Across the test ranges $10^{13.7}\leq M_{\rm
  vir}\leq10^{15.3}$\hMsun and $10^{1.7}\leq r_{\rm
  s}\leq10^{2.7}$\hkpc, all configurations yield nearly unbiased
absolute median residuals (within 0.01~dex).  For the halo mass, adding
richness narrows the residual distribution, reducing the standard
deviation from 0.133 to 0.122~dex for the all sample, and from 0.124 to 0.111~dex for the relaxed sample. For the scale radius, restricting the training to relaxed clusters improves the performance more than adding richness. The standard deviation decreases from 0.180 to 0.154~dex for \cnnb\ and from 0.175 to 0.148~dex for \cnnr, while the inclusion of richness yields only a modest improvement of 0.005~dex. These results demonstrate that machine learning is a powerful tool to infer the mass and internal mass distribution of clusters, providing a new window for cosmological inferences and understanding galaxy formation processes.
\end{abstract}

\section{Introduction} \label{sec:sec1}

Galaxy clusters are the most luminous objects in the Universe and
serve as cosmological probes. The abundance of clusters is sensitive
to underlying cosmological parameters such as the cosmic matter
density ($\Omega_{\rm m}$) and the root-mean-square mass fluctuation on 8\hMpc scale 
\citep[$\sigma_8$, e.g.,][]{Wang1998, Allen2011}.

The challenge in constraining cosmological parameters utilizing cluster
abundances is to robustly estimate cluster masses \citep{Allen2011}.  Cluster mass is not
a directly observed quantity; therefore, several mass proxies spanning
a variety of wavelengths have been used to indirectly estimate mass,
including the X-ray luminosity from the hot intracluster medium
\citep[e.g.,][]{Pratt2009, Vikhlinin2009} and the weak gravitational lensing  
\citep[e.g.,][]{Wilson96, Holhjem09, Hoekstra2013}.  The
properties of member galaxies have also been used, including their
velocity dispersion \citep[e.g.,][]{Biviano2006, Evrard2008}, as well as 
richness, which is the number of member galaxies in a cluster 
\citep{Rozo2009, Rykoff2012, Simet2017, 2023ApJ...955...26A}.
All of these quantities are tightly correlated with the cluster's total mass.

Not only the cluster mass, but also the internal mass distribution of a cluster's dark matter halo
contains a wealth of information about its assembly history and the underlying
cosmological parameters.  Cosmological simulations predict that the spherically
averaged density profile of halos is well fitted by the
Navarro--Frenk--White (NFW) profile \citep{n1996,n1997} described by
\begin{eqnarray}
\rho_{\mathrm{NFW}}(r) = \frac{\rho_{\mathrm{s}}}{\frac{r}{r_{\mathrm{s}}}\left(1 + \frac{r}{r_{\mathrm{s}}}\right)^{2}} \label{eq:rs1},
\end{eqnarray}
where, $r$ is a distance from the halo center, $\rho_{\mathrm{s}}$ is
a characteristic density, and \rs is a scale radius.  The
concentration parameter is defined as $c = R_{\mathrm{vir}} /
r_{\mathrm{s}}$, where $R_{\mathrm{vir}}$ is the halo virial radius, and
depends on the halo mass and redshift 
\citep[e.g.,][]{Bullock2001, Prada2012, Klypin2016, Diemer2019, Ishiyama2021}. 
The concentration is a key
parameter of halos through correlations with several internal halo
properties such as sphericity, relaxation state, and subhalo abundance
\citep[e.g.,][]{Ishiyama2009, Jeeson-Danie2011, Skibba2011}, 
and halo assembly history \citep[e.g.,][]{Wechsler2002}.
The amplitude of halo spatial clustering primarily depends on the halo mass
\citep{Kaiser1984,Mo1996} and also depends on secondary properties of
a halo such as the concentration \citep{Gao2005,Wechsler2006,Gao2007}, which is
referred to {\it assembly bias}.
This assembly bias is crucial for cosmological inferences; therefore, it serves as a cosmological probe by itself, 
challenging us to detect it through observations. 

\citet{Miyatake2016} and \citet{Surhud2016} reported the detection of assembly
bias using the redMaPper galaxy cluster catalog \citep{Rykoff2014}.
However, follow-up studies showed that the detected signal was larger
than the prediction by cosmological simulations probably due to
interloper galaxies along the line of sight direction of a cluster
\citep{Busch2017, Zu2017, Sunayama2019}, which caused systematic errors in the
estimation of concentration.  Therefore, it is necessary to robustly
estimate cluster halo mass and concentration. 
However, it is particularly challenging to observationally infer 
cluster concentration with high accuracy due to projection effects, halo triaxiality, 
and selection biases \citep[e.g.,][]{Oguri2012, Hoekstra2013, Pratt2019}.

A new approach utilizing rapidly developing machine learning (ML) is 
a potential candidate to shed light on this issue 
\citep[e.g.,][]{Ntampaka2015, Ntampaka2016, h2019, Ho2021, Soltis2025}. 
\citet{h2019} studied the ability of convolutional neural networks (CNNs)
to infer cluster mass using two observables of member galaxies:
the projected distance and the line-of-sight velocity. 
They showed that CNNs could be superior to a classical estimator using
the scaling relation between the cluster mass and velocity dispersion of
member galaxies. Although this approach stands out under
a fixed cosmological model, it enables us to take observational
systematics into account.
Later, this approach was extended using Bayesian neural networks to predict 
the mass posterior \citep{Ho2021}.

In this study, we extend the model by \citet{h2019} and investigate
the ability of CNNs for the inference of both cluster mass and
concentration using a large dataset of mock galaxies, the
Uchuu-UniverseMachine galaxy catalog \citep[Uchuu-UM:][]{uchuu-um2023}
based on the Uchuu cosmological $N$-body simulations
\citep{Ishiyama2021}.  Compared to the MultiDark Planck2 simulation
\citep{Klypin2016} used in \citet{h2019}, the Uchuu simulation is
superior in terms of an eight-times larger volume and a five-times higher
mass resolution, providing a better training dataset for ML.

The structure of this paper is as follows. 
\S~\ref{sec:sec2} describes the procedure for creating the dataset using simulated galaxies.
\S~\ref{sec:sec3} presents an explanation of the CNN model. 
In \S~\ref{sec:sec4}, the performance of the model using the CNN is evaluated.
Finally, \S~\ref{sec:sec5} concludes the paper.
Throughout this paper, we use the units of \hMsun and \Msun 
for halo and stellar mass of a galaxy, respectively, 
and the term ``$\log$'' as ``$\log_{10}$'' unless otherwise stated.

\section{Dataset for Machine Learning}\label{sec:sec2}

In this section, we describe the details of our dataset for ML.  We
perform mock observations to galaxy clusters and galaxies
extracted from the Uchuu-UM galaxy catalog
\citep{uchuu-um2023}, which was constructed by applying an empirical
galaxy formation model,
\textsc{UniverseMachine}~\footnote{\url{https://bitbucket.org/pbehroozi/universemachine/src/main/}}
\citep{Behroozi2019}, to one of the largest cosmological $N$-body
simulations, Uchuu \citep{Ishiyama2021}.  The Uchuu halo catalogs and the
Uchuu-UM galaxy catalogs are publicly available on the \SU site~\footnote{\urluchuu}.
For the mock observations, we follow the procedure presented in \citet{h2019}.

\subsection{Cluster Catalog}\label{sec:sec21}
The Uchuu simulation \citep{Ishiyama2021} consists of a box of
$2.0\,h^{-1}\,\mathrm{Gpc}$ on a side with $12800^{3}$ particles, 
corresponding to the mass resolution of $3.27\times 10^{8} \,h^{-1} \, \mathrm{M_{\odot}}$. 
The cosmological parameters of this simulation are based on the 
Planck 2016 cosmology \citep{Planck2016}, namely 
$\Omega_{\rm m} =0.3089$, $\Omega_{\rm b} = 0.0486$, $h=0.6774$, $n_{\rm s} = 0.9667$, 
$\Omega_\Lambda = 0.6911$, and $\sigma_{8} = 0.8159$.  
Halos and subhalos were identified using \textsc{Rockstar} phase space halo/subhalo 
finder~\footnote{\url{https://bitbucket.org/gfcstanford/rockstar/}
An extension parallelized by Message Passing Interface (MPI), \textsc{MPI-Rockstar} 
 is also available at \urlmrockstar \citep{Tokuue2024}.} \citep{Behroozi2013}.
Merger trees were constructed using the 
\textsc{consistent trees merger tree
  code}~\footnote{\url{https://bitbucket.org/pbehroozi/consistent-trees/}}
\citep{Behroozi2013b}.

Mock galaxies were modelled on the Uchuu merger trees using an
empirical galaxy formation model, \textsc{UniverseMachine}
\citep{Behroozi2019}. The basic statistics of the Uchuu-UM catalog are
presented in \citet{uchuu-um2023}.  In this study, we extract galaxy
clusters with $M_{\mathrm{vir}} \geq 10^{13.5} \, h^{-1}\,
\mathrm{M_{\odot}}$ and galaxies with $M_{\rm{stellar}} \geq 10^{9.5}
\, \mathrm{M_{\odot}}$ from the Uchuu-UM at $z = 0.093$, where \mvir
and $M_{\rm{stellar}}$ are the halo virial mass and galaxy stellar
mass, respectively. The virial overdensity is based on the spherical
collapse model \citep{Bryan1998}.

The definition of concentration is the virial radius divided by the
scale radius.
\textsc{Rockstar} calculates the scale radius by fitting
the spherically averaged density profile with the NFW profile. 
Besides, \textsc{Rockstar} provides an alternative scale radius obtained by numerically solving 
\begin{eqnarray}
V_{\rm vir} = \left( \frac{GM_{\rm vir}}{R_{\rm vir}} \right)^{1/2}, {\rm and} \quad \frac{V_{\rm max}}{V_{\rm vir}} = \left( \frac{0.216c}{f(c)} \right)^{1/2}\label{eq:vvir}, 
\end{eqnarray}
where $f(c)=\ln(1+c)-c/(1+c)$ and $V_{\rm max}$ is the maximum
circular velocity \citep{Klypin2011, Prada2012}.
We choose the latter scale radius as an output of the ML 
to avoid the ill-defined scale radius due to the fitting. 

We naively expect that the performance of the ML model should be
improved if we use only dynamically relaxed clusters rather than all
clusters. Therefore, we construct two cluster catalogs: (i) {\it all
  sample}, which contains all clusters that satisfy the aforementioned
conditions; and (ii) {\it relaxed sample}, which contains only
dynamically relaxed clusters among them; and compare the performance
of ML for these two catalogs.  This treatment puts the stark contrast
with the study by \citet{h2019}, which used only all sample.  We use
the following three conditions proposed by \citet{Klypin2016} to select
dynamically relaxed clusters.
\begin{eqnarray} 
  \frac{2K}{|W|} < 1.5, \quad X_{\mathrm{off}} < 0.07, \quad \lambda < 0.07, 
\label{eq:k2016} 
\end{eqnarray}
where, $K$, $W$, and $\lambda$ are the kinetic, potential energies, and the spin parameter of a halo, respectively.
The offset parameter $X_{\mathrm{off}}$ is defined as the distance between the center
of a halo and its center of mass.

\begin{figure*}
\includegraphics[width=0.48\linewidth]{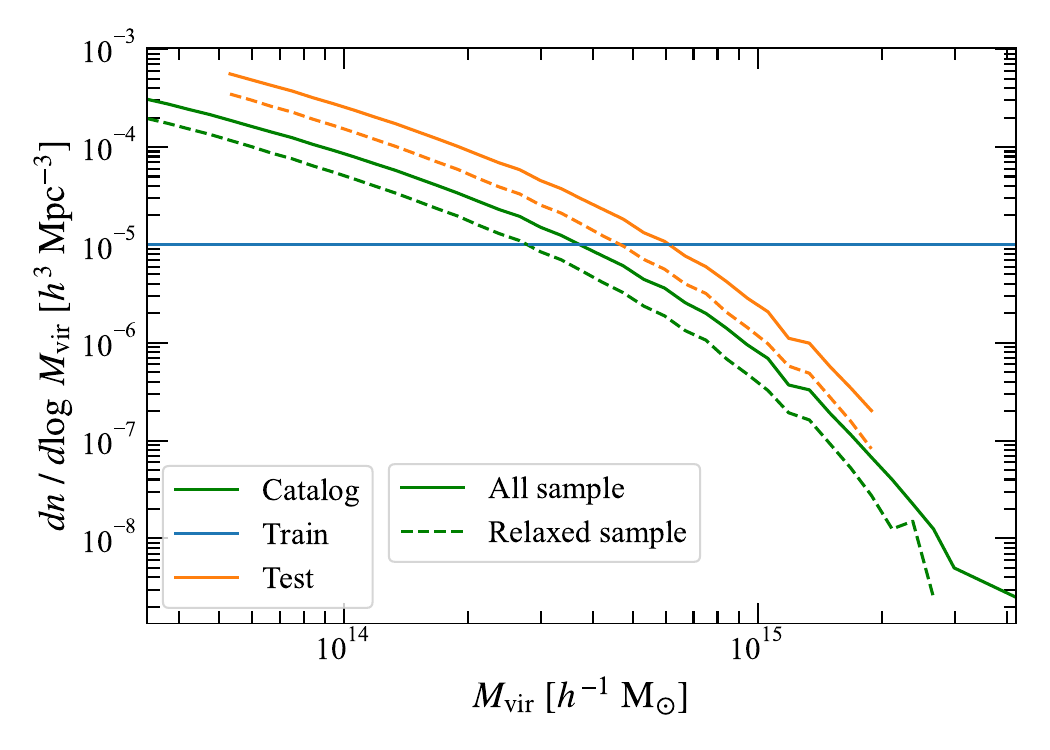}
\includegraphics[width=0.48\linewidth]{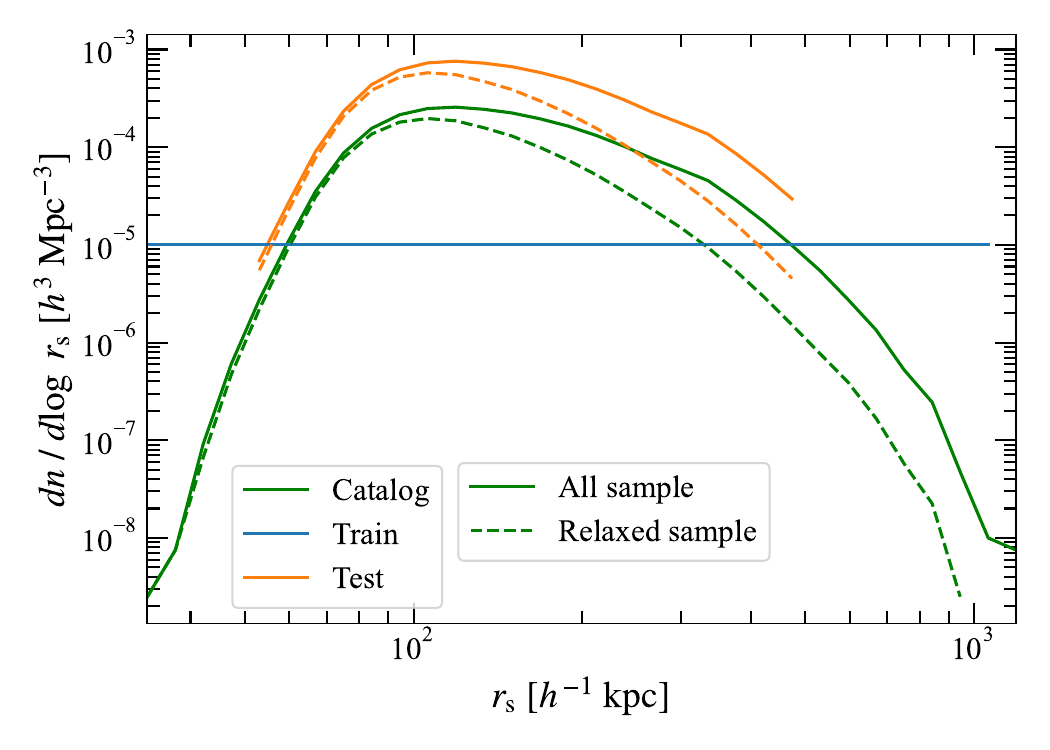}
\caption{
Number densities of clusters as functions of virial mass (left) and scale radius (right), shown per logarithmic interval (per dex) and in units of \hMpccube. 
Solid and dashed curves correspond to the full and relaxed samples, respectively. 
Green, cyan, and orange curves show the original Uchuu-UM catalog, the training dataset, and the test dataset, respectively. 
For the training dataset, the curves for the full and relaxed samples nearly overlap in both panels.
}
\label{fig:cluster_func}
\end{figure*}

\subsection{Mock Observation}\label{sec:sec22}

We perform realistic mock observations for the galaxy clusters 
to use the cluster observables as input data for our ML model. 
Following the procedure described in the Appendix of \citet{h2019},
for each galaxy of a cluster,
we calculate the projected radial distance onto the plane of the
sky, \rproj, based on the Euclidean distance between the centers of the cluster and galaxy, 
and the line-of-sight (LOS) relative velocity between them, \vlos.
To determine \vlos, we place the center of cluster at $z=0.093$ and 
calculate the distance from an observer at $z=0$ to it. 
Then, we calculate the redshift of the galaxy based on the LOS distance 
from the cluster. We add corresponding Hubble velocities to both LOS 
comoving peculiar velocities, and use the difference of them as \vlos.
Finally, we calculate a two-dimensional joint probability distribution function (PDF) of \rproj and \vlos for each cluster 
and use it as input data for our ML model as described in \S~\ref{sec:sec3}.

In realistic observations, it is difficult to avoid the contamination
of interloper galaxies due to projection effects.  To mimic observations,
instead of using true membership based on the hierarchy of halos and subhalos,  
we consider galaxies with \rproj$\leq 1.6$\hMpc and $|v_{\rm los}| \leq
2200$\kms around each cluster and regard them as member galaxies.
As a result, member galaxies distribute in a cylindrical shape. 
We limit the determination of member galaxies within a 112\hMpc 
cube centered on the cluster to reduce calculation costs by following \citet{h2019}.

To avoid introducing significant bias during training, we construct a training dataset of mock clusters such that the distributions of mass and scale radius are approximately flat (equal counts per bin) across the full training ranges. Figure~\ref{fig:cluster_func} shows the cluster number densities as functions of virial mass and scale radius, expressed per logarithmic bin (per dex) and in units of \hMpccube, demonstrating highly non-uniform distributions. We flatten these distributions by downsampling clusters in high number density bins and upsampling clusters in low number density bins, using a target threshold of $10^{-5}\,$\hMpccube\ for both the virial mass and scale radius distributions.
The downsampling is performed by randomly selecting clusters.
The upsampling is performed by placing multiple observational points around each cluster. 
These points are generated using randomly rotated Fibonacci lattices \citep{g2010}, which yield an approximately uniform distribution on the surface of a sphere. 
For each cluster, the sphere radius is set to the comoving distance between the observational point and the cluster. In contrast, we construct the test dataset by performing three orthogonal projections for each cluster, independent of the virial mass and scale radius distributions. 
The number densities of the training and test datasets are also shown in Figure~\ref{fig:cluster_func}.

The number of member galaxies of clusters (richness) strongly
correlates with the clusters' total masses. This correlation is known as the
mass-richness relation \citep{Rozo2009, Rykoff2012, Simet2017, 2023ApJ...955...26A}.  Based on
this fact, we also try to improve the performance of mass and scale radius 
inference by incorporating richness as an additional input to the ML model. In this study, we define the richness as the number of member galaxies
including interlopers instead of the number of true member galaxies to mimic the realistic observational conditions.  We exclude clusters with richness fewer than 10 to prevent a significant accuracy loss of ML  by following \citet{h2019}. Finally, the total number of clusters of the full and relaxed samples extracted from the catalog are 935,144 and 576,076, respectively.
The total number of true member galaxies contained within the virial radius of both cluster samples are 25,091,876 and 15,073,204, respectively;
however, the contamination of interloper galaxies increases the richness used in the ML models. 

\begin{figure*}
\includegraphics[width=\linewidth]{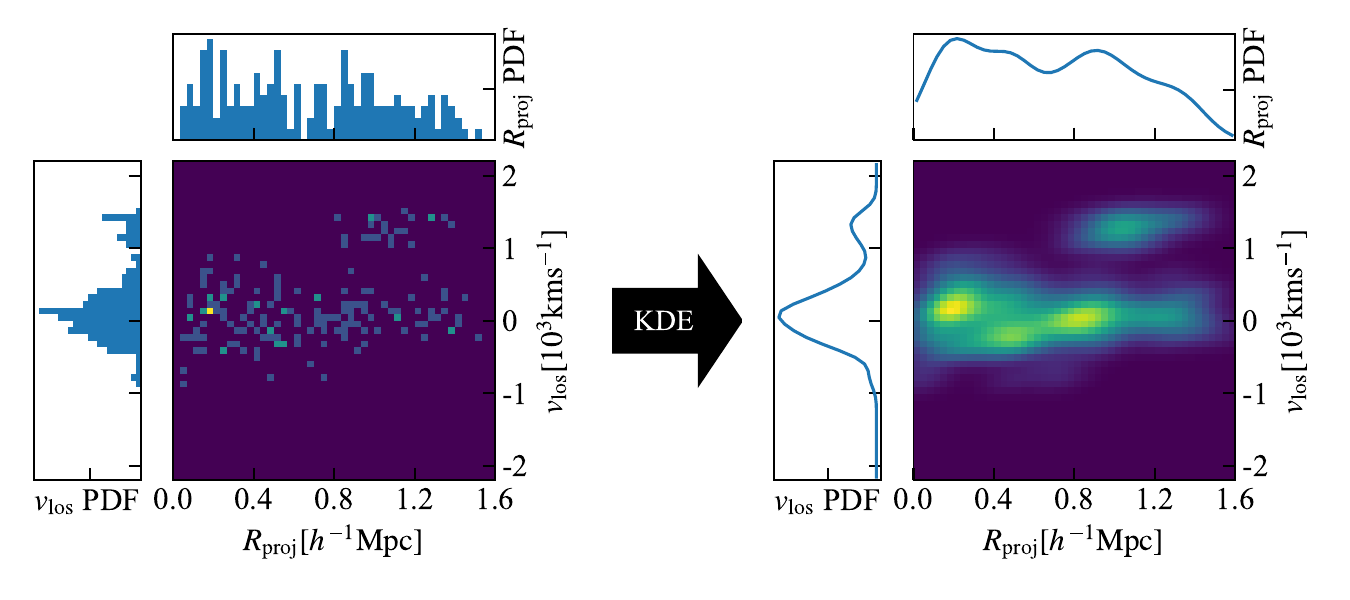}
\vspace{-0.5cm}
\caption{
Heat-map images of the two-dimensional joint PDF in the \rproj--\vlos plane for a representative cluster, shown before (left) and after (right) applying KDE. The example is constructed from 169 galaxies. Both \rproj and \vlos are discretized into 48 bins. The left panel shows the pre-KDE heat-map constructed from the one-dimensional PDFs of \rproj\ and \vlos\ that directly reflect the observations described in \S~\ref{sec:sec2}. The right panel shows the corresponding KDE-smoothed density field adopted as the CNN input in this study.
}
\label{fig:heatmap}
\end{figure*}


\section{CNN Architecture}\label{sec:sec3}

In this section, we describe the details of our ML model based on the
cluster observables for the inference of cluster mass and scale
radius.  We construct CNNs using the
TensorFlow API version 2.19.0 \footnote{\url{https://www.tensorflow.org/}} and Keras libraries version 3.9.2\footnote{\url{https://keras.io/}}.

\subsection{Input Data}\label{sec:sec31}

The procedure used to construct the input images for the CNNs is largely based on \citet{h2019}. 
For each cluster, we construct a two-dimensional joint PDF in the \rproj--\vlos plane for member galaxies (including interlopers). 
We discretize this PDF into a $48\times48$ grid with equal-size bins. Each PDF is treated as a $48\times48$ pixel image for our CNN models. 
Figure~\ref{fig:heatmap} shows an example PDF. Pixels containing galaxies are sparse because the number of member galaxies (including interlopers) in each image is typically smaller than the number of pixels. To mitigate biases that could arise from this sparsity, we smooth each PDF into a continuous density field using kernel density estimation (KDE), defined as
\begin{eqnarray} 
\hat{f}(x) = \frac{1}{nw}\sum_{i=1}^{n}K\left( \frac{x-x_{i}}{w} \right), \label{eq:1}
\end{eqnarray}
where $K$ and $w$ are the kernel function and bandwidth, respectively. Given a set of samples $\{x_i\}$ drawn from a variable $x$, the KDE estimate of the PDF is $\hat{f}(x)$.
We use the Gaussian KDE for the kernel function, and set the
kernel bandwidth to $w=0.25\sigma$, where $\sigma$ denotes the sample standard deviation, following Scott's rule \citep{sc2015}. 
Figure~\ref{fig:heatmap} also illustrates the KDE inputs provided to the network. 

\begin{figure*}
\includegraphics[width=\linewidth]{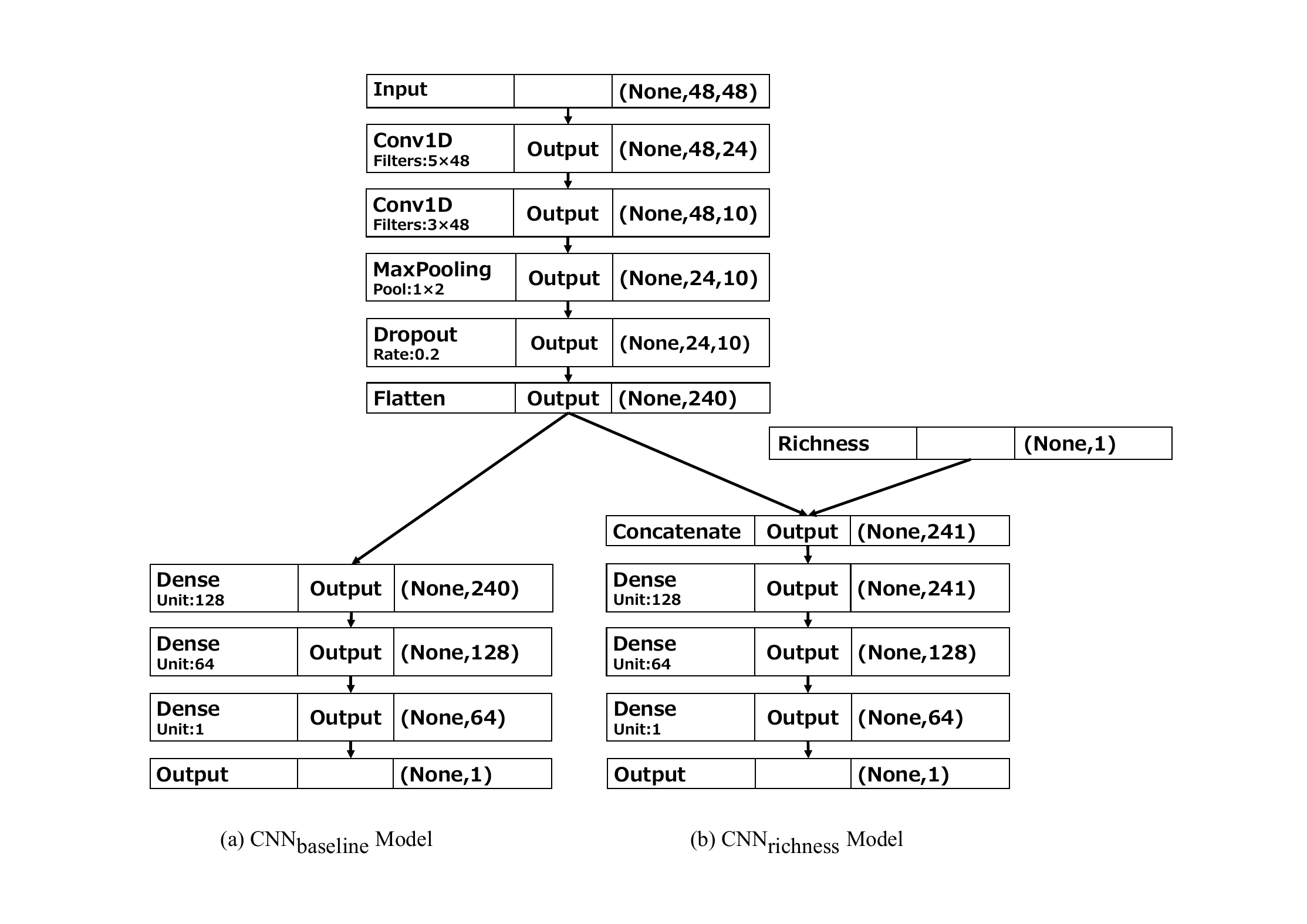}
\vspace{-1.0cm}
\caption{
CNN architecture created using Keras library on TensorFlow. The
input consists of $48 \times 48$ pixel image described in \S~\ref{sec:sec2}
and~\ref{sec:sec31}. The output shape of each layer is indicated in
parentheses. The CNN architecture shown in (a) follows the same layer
configuration as Ho et al.\ (2019), whereas (b) illustrates a proposed
network in the present study. The Adam optimizer was used with the
Keras/TensorFlow default settings, except that the learning rate and the decay were
set to $10^{-6}$ and $0.999$, respectively.
}
\label{fig:CNN}
\end{figure*}

\subsection{CNN layers}

In this work, we construct two different CNNs: (i) \cnnb, which adopts the same
architecture used in \citet{h2019}; and (ii) \cnnr, which is an
extension of \cnnb by adding an additional input layer for
richness.  Figure~\ref{fig:CNN} illustrates the schematic diagrams of
CNN architecture, and their details are given as follows.

\cnnb\ consists of two convolutional layers, one max-pooling layer, one dropout layer, one flatten layer, three dense layers, and a single linear output layer. In the first convolutional layer, a $5\times48$ filter is applied to the $48\times48$ input images constructed following the procedure described in \S~\ref{sec:sec31}, producing 24 feature maps (each of width 48). A second convolutional layer with a $3\times48$ filter is then applied, producing 10 feature maps, also with width 48. Next, a max-pooling layer with a $1\times2$ kernel and a stride of one downsamples the feature maps, followed by a dropout layer with a rate of 0.2 to reduce overfitting. The resulting two-dimensional feature maps are flattened into one-dimensional vectors of length 240, which are then passed through three dense layers to produce the final output.

In \cnnr, the richness is added to the tail of the flattened vector of size 240; therefore, the input vector size for the first Dense layer is 241.  This is the only difference between \cnnb and \cnnr.

The output $y$ of CNNs is constrained to the range $0 \le y \le 1$ 
and corresponds to linear scaling of logarithmic cluster mass and scale radius as
\begin{eqnarray}
  \log{\qty[Y_{\rm pred}]} = \log{\qty[Y_{\rm min}]}
  + y \log{\qty[ \frac{Y_{\rm max}}{Y_{\rm min}}]}, 
\label{eq:Y} 
\end{eqnarray}
where, $Y_{\rm min}$ and $Y_{\rm max}$ are the minimum and maximum
true values in the training dataset (see \S~\ref{sec:train}), respectively, namely those of
virial mass or scale radius, and $Y_{\rm pred}$ is the value predicted by CNNs.
This normalization improves the training stability and ensures that
the predicted values fall within a physically meaningful range.
However, since $Y_{\rm min}$ and $Y_{\rm max}$ are determined based on
values within the training range, input parameters outside this range
may result in values of $y < 0$ or $y > 1$.

\subsection{Training} \label{sec:train}

\begin{figure}
\includegraphics[width=\linewidth]{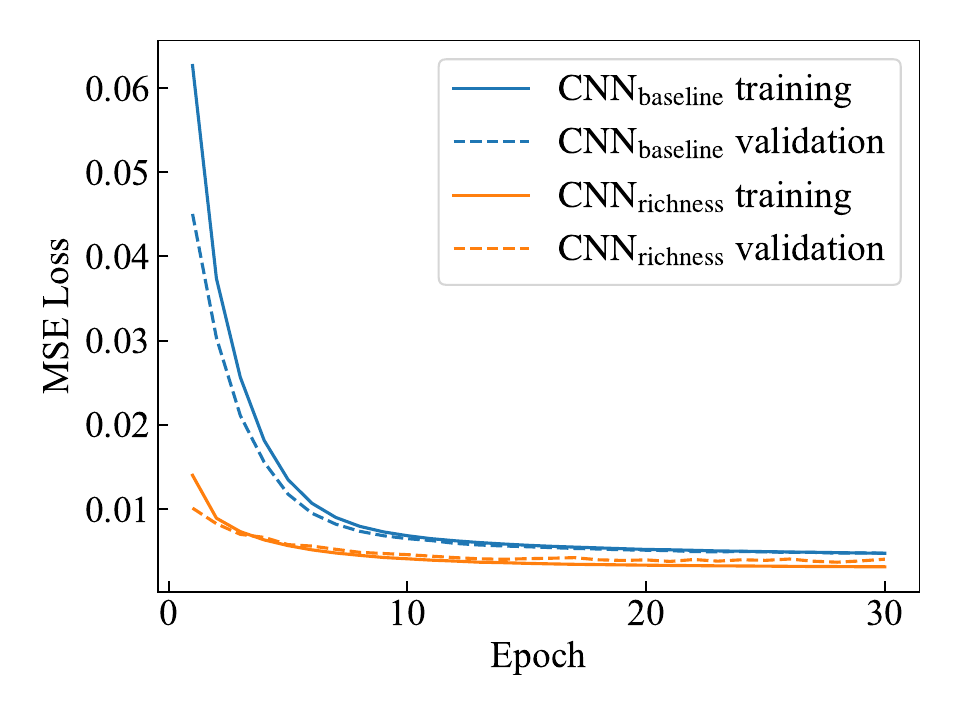}
\caption{
The evolution of MSE loss for a single fold in each of \cnnb and \cnnr
during the training of \mvir.  These folds have the smallest final MSE
loss among the 10 folds in each CNN architecture at the 30th epoch.  In
this training, the MSE loss of all folds nearly converges before 30
epochs.
}
\label{fig:loss}
\end{figure}

To optimize the output value $y$ during the training, we use the mean
squared error (MSE) with the kernel regularization with $L2=0.1$ as
the loss function and the ReLU activation function \citep{ReLU} for
both the convolutional and dense layers.  The Adam optimizer
\citep{Kingma2014AdamAM} is used for parameter optimization.  
To stabilize the training, we use
the Keras/TensorFlow default settings, except that the learning
rate and the decay were set to $10^{-6}$ and $0.999$, respectively.
The batch size and the maximum epochs are 100 and 30, respectively.

We train the CNN models using 10-fold cross-validation based on 10
subsets of the training data. The training and the validation ranges
are the same, $10^{13.5} \leq M_{\rm vir} \leq 10^{15.7}$\hMsun and
$10^{1.5} \leq r_{\rm s} \leq 10^{3.1}$\hkpc.  In the $k$-th iteration
of cross-validation, the $k$-th subset is used as the validation data,
whereas the remaining nine subsets are used as the training data.  We
divide the training data into 10 disjoint subsets such that the same cluster
with different LOS is assigned to the same subset, ensuring the same
cluster is never used in both training and validation data.  To evenly
assign unique clusters to each subset, clusters are first binned in
intervals of 0.05~dex based on the mass or the scale radius, then
clusters are randomly shuffled and partitioned into 10 equal-sized
subsets.  This procedure avoids unevenly assigning clusters at
the high mass or the high/low scale radius into each subset because
such clusters are less abundant, as seen in Figure~\ref{fig:cluster_func}.

Figure~\ref{fig:loss} shows the MSE loss for a single fold chosen from
10-fold cross-validation for training \mvir using \cnnb and \cnnr
architectures.  The MSEs rapidly decrease at initial epochs and
gradually decrease until reaching a nearly constant value regardless
of the architecture.  
The change in both MSE loss values after the 10th epochs
is small.  To monitor the loss function and prevent overfitting, we
use the callback API of the \texttt{EarlyStopping} class provided by
Keras library with $\mathrm{monitor}=\mathrm{'val\_loss'}$ and keep
all other parameters at their defaults.  The \texttt{EarlyStopping}
API monitors the loss function every epoch and terminates training
when the MSE of loss function converges.  The number of epochs was four for the
fold that was stopped earliest, where the MSE loss was below $\sim$0.01. 
The difference in MSE lost between training and validation is small, 
ensuring little effect of overfitting. 

After the 10-fold cross-validation, each derived CNN independently
infers the mass and the scale radius of clusters in the test dataset.
The test ranges are $10^{13.7} \leq M_{\rm vir} \leq 10^{15.3}$\hMsun and  
$10^{1.7} \leq r_{\rm s} \leq 10^{2.7}$\hkpc.
The final predicted value of each cluster is computed by averaging the
inferences obtained from the 10 folds.

\begin{figure}
\includegraphics[width=1.0\linewidth]{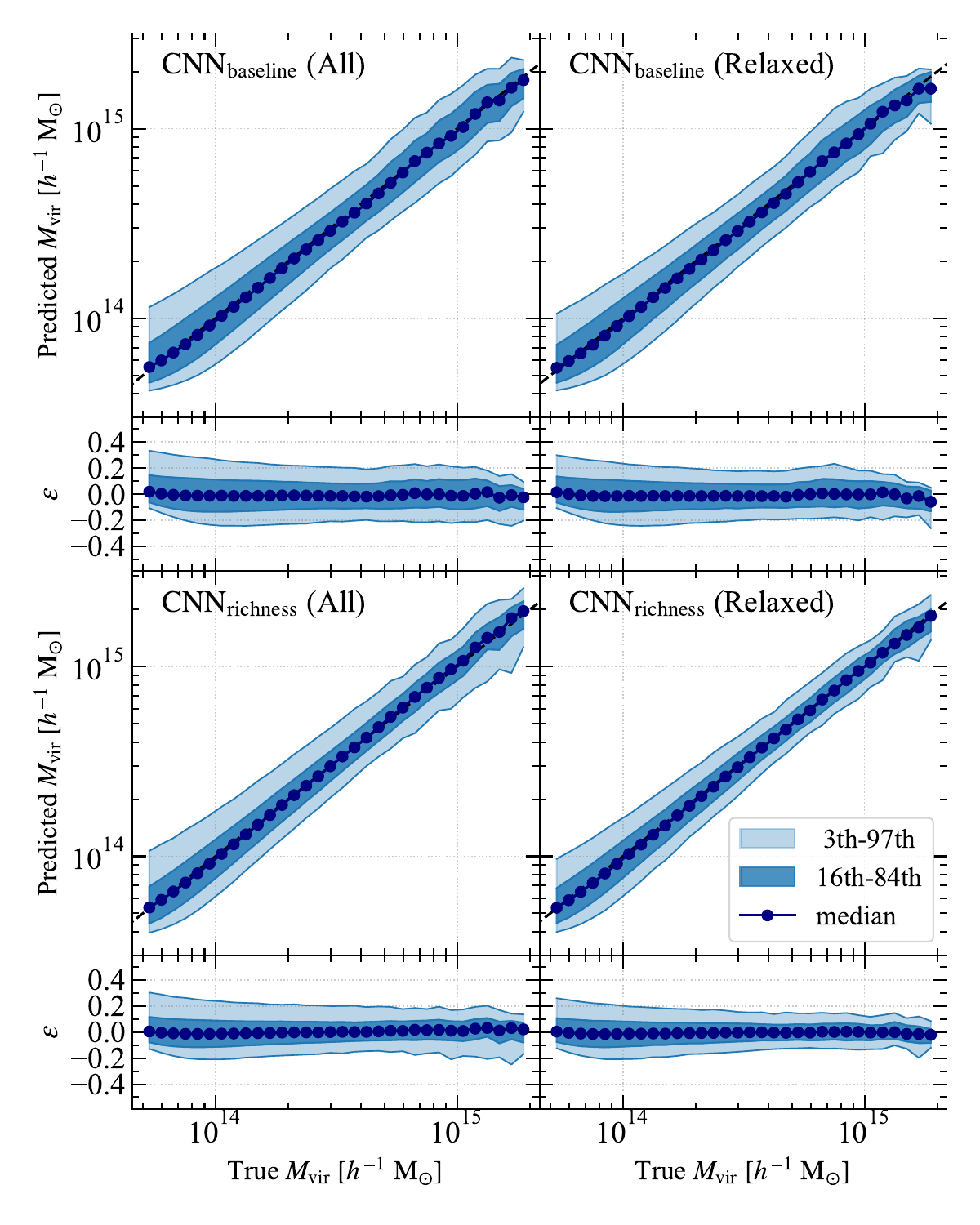}
\caption{Virial mass recovery: Top and bottom panels in each of the four plots show distributions of
predicted versus true virial mass and their residuals $\epsilon$
defined in Equation \eqref{eq:epsilon}, respectively.  Circles show
the median of predicted values in each true mass bin.  Thick and thin
shaded areas show the 16th-84th and 3rd-97th percentile scatter
ranges, respectively.  Dashed lines indicate equality between predicted and true values.}
\label{fig:mvir_pred}
\end{figure}

\begin{figure}
\includegraphics[width=1.0\linewidth]{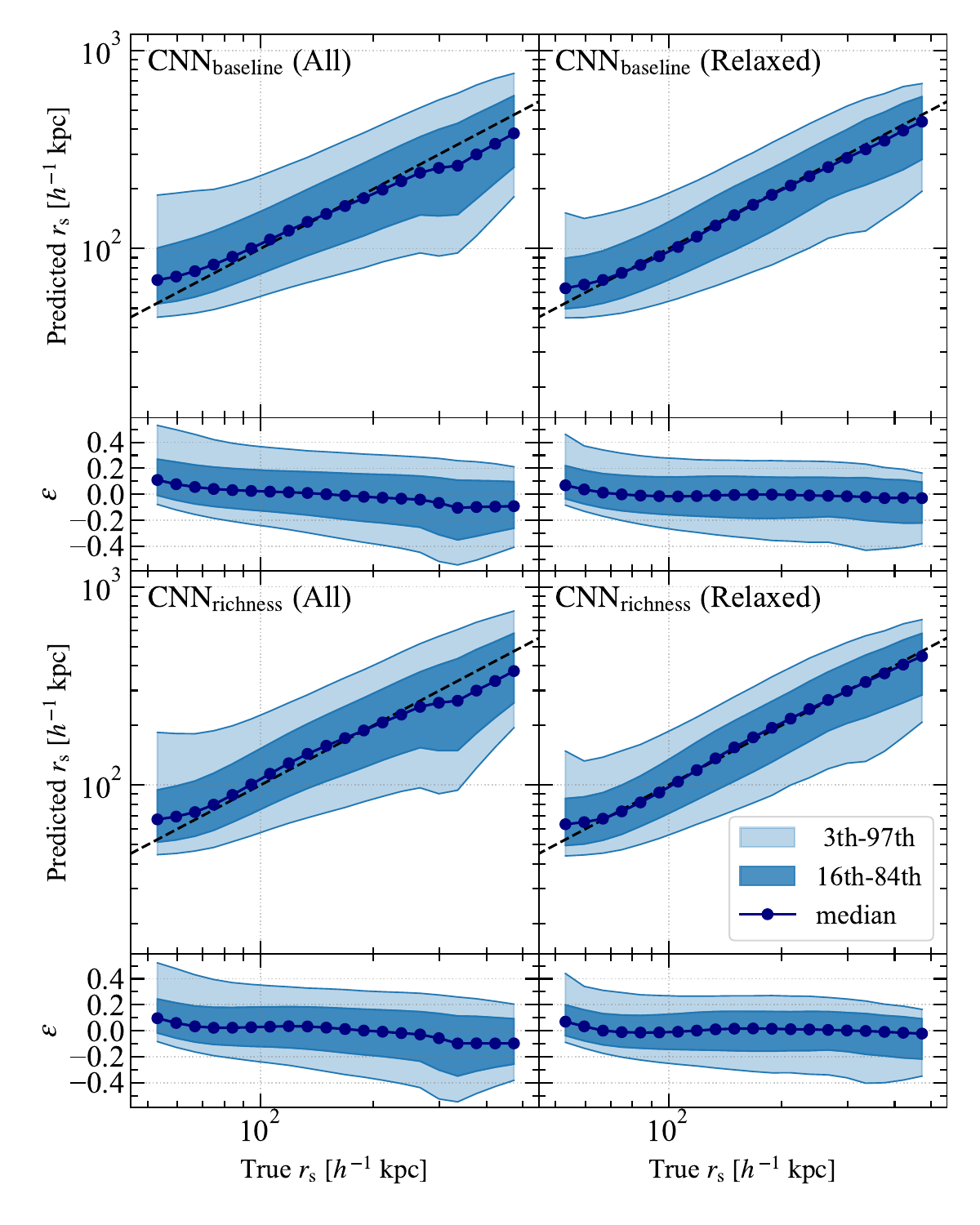}
\caption{Scale radius recovery:
Top and bottom panels in each of the four plots show distributions of
predicted versus true scale radius and their residuals $\epsilon$
defined in Equation \eqref{eq:epsilon}, respectively.  Circles show
the median of predicted values in each true mass bin.  Thick and thin
shaded areas show the 16th-84th and 3rd-97th percentile scatter
ranges, respectively. Dashed lines indicate equality between predicted and true values.}
\label{fig:rs_pred}
\end{figure}

\begin{deluxetable*}{ccc crrrrcrrrr}
\tablenum{1}
\tablecaption{Summary of CNN predictions, namely
median $\tilde{\epsilon}$ (dex), 16th-84th percentile range
$\delta\epsilon$ (dex), standard deviation scatter $\sigma_{\epsilon}$ (dex),
skewness $\gamma$, and excess kurtosis $\kappa$ calculated from
$\epsilon$ in Equation \eqref{eq:epsilon}.  
\label{table:pred_summary}}
\tablewidth{0pt}
\tablehead{
\colhead{Inference} & \colhead{CNN} & \colhead{Sample} & &
\colhead{$\tilde\epsilon\pm\delta\epsilon$} & \colhead{$\sigma_{\epsilon}$} & \colhead{$\gamma$} & \colhead{$\kappa$}
}
\startdata
\multirow{4}{*}{$\rm{CNN}_{\rm{baseline}}$} & \multirow{2}{*}{All} & $M_{\rm{vir}}$ && $-0.004^{+0.122}_{-0.104}$ & 0.133& 1.279& 5.785\\
 &  & $r_{\rm{s}}$ && $0.002^{+0.168}_{-0.174}$ & 0.180& 0.088& 0.759\\
\cline{2-8}
 & \multirow{2}{*}{Relaxed} &$M_{\rm{vir}}$ && $-0.006^{+0.116}_{-0.101}$ & 0.124& 1.058& 4.773\\
 & &$r_{\rm{s}}$ && $-0.009^{+0.145}_{-0.153}$ & 0.154& 0.181& 0.912\\
\hline
\multirow{4}{*}{$\rm{CNN}_{\rm{richness}}$} &\multirow{2}{*}{All} & $M_{\rm{vir}}$ && $-0.006^{+0.100}_{-0.092}$ & 0.122& 1.891& 10.740\\
 & & $r_{\rm{s}}$ && $0.013^{+0.159}_{-0.173}$ & 0.175& -0.003& 0.882\\
\cline{2-8}
 &\multirow{2}{*}{Relaxed} &$M_{\rm{vir}}$ && $-0.008^{+0.094}_{-0.089}$ & 0.111& 1.581& 9.166\\
 & &$r_{\rm{s}}$ &&  $0.000^{+0.137}_{-0.144}$ & 0.148& 0.230& 1.247\\
\enddata
\end{deluxetable*}

\begin{figure}
\includegraphics[width=\linewidth]{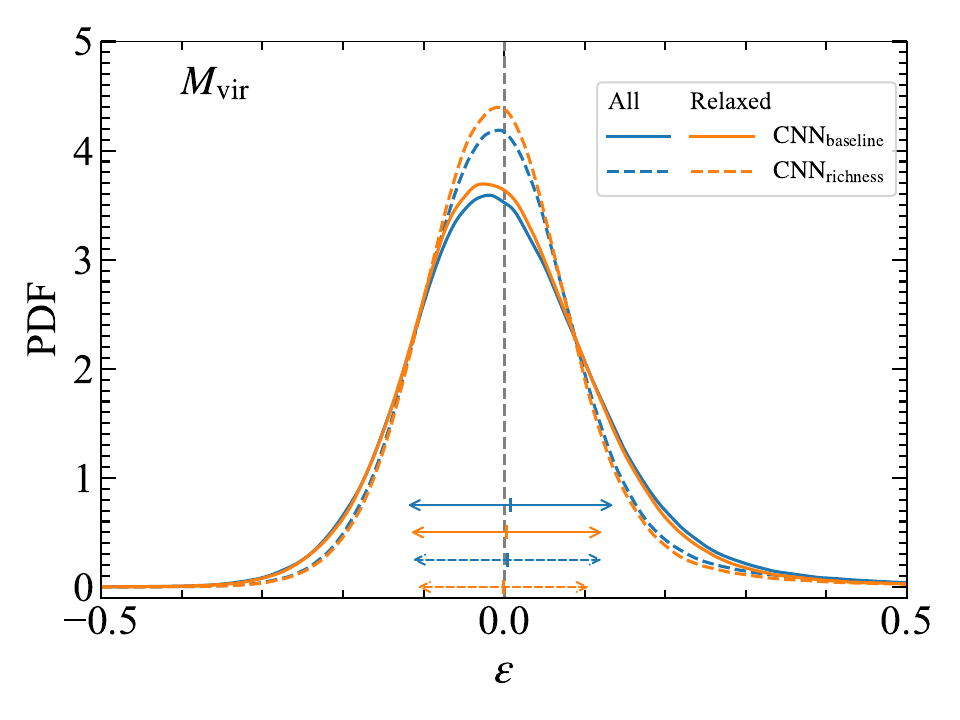}
\includegraphics[width=\linewidth]{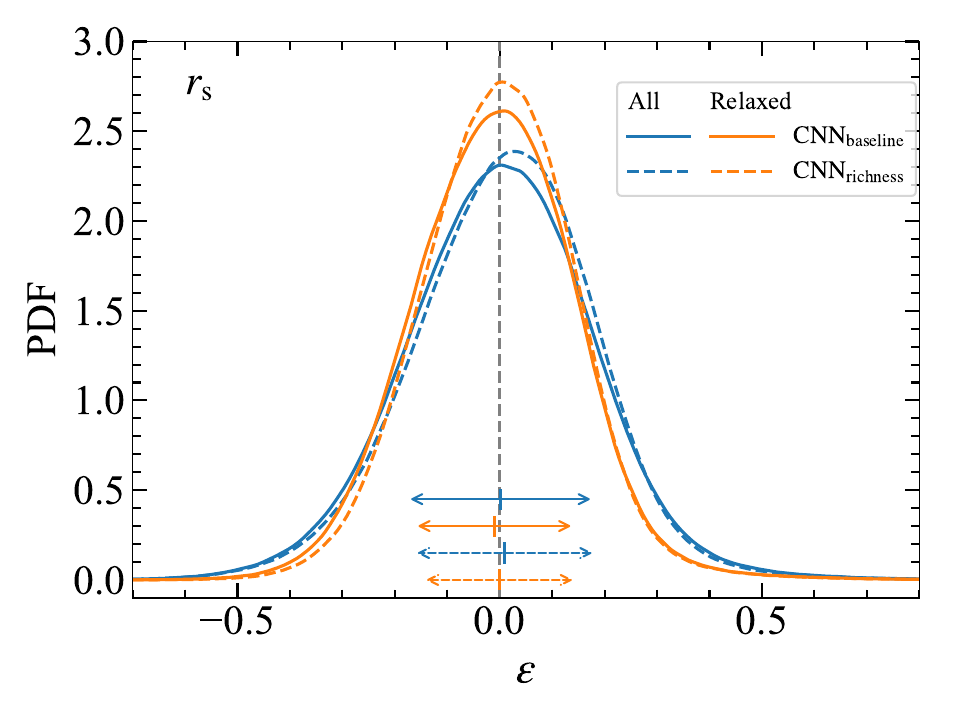}
\caption{
Probability distribution function (PDF) of prediction residuals 
$\epsilon$ for the virial mass (top) and the scale radius (bottom). 
Arrows and their centers indicated by vertical short lines represent $\pm1$ standard deviation spreads 
and the average of $\epsilon$, respectively. 
}
\label{fig:residual_distribution}
\end{figure}

\section{Results}\label{sec:sec4}

In this section, we present the performance of our CNN models 
for inferring halo mass and scale radius.
To quantify model performance, we use the logarithmic 
residual defined as
\begin{eqnarray}
\epsilon = \log{\qty[ \frac{Y_{\rm pred}}{Y_{\rm true}}]},
\label{eq:epsilon}
\end{eqnarray}
where, $Y_{\rm {true}}$ and $Y_{\rm {pred}}$ are the true and
predicted values, respectively, namely those of virial mass or scale
radius.  The prediction results are summarized in
Figures~\ref{fig:mvir_pred}--\ref{fig:residual_distribution} and
Table~\ref{table:pred_summary}.  Figures~\ref{fig:mvir_pred} and~\ref{fig:rs_pred} 
compare the distributions of predicted versus true
virial mass and scale radius for the four CNN models,
respectively.  Probability distribution functions of prediction
residuals are illustrated in Figure~\ref{fig:residual_distribution}.
Table~\ref{table:pred_summary} presents statistical results of the
distribution of the prediction residual.

\subsection{Inference of the halo mass}\label{sec:results:mvir}

In all CNN models, the median predicted mass is unbiased and reproduces well the true mass over the mass range probed by the test set (Figure~\ref{fig:mvir_pred}). The 16th--84th and 3rd--97th percentile scatter bands are well centered and narrow, as summarized in Table~\ref{table:pred_summary}.  In all cases, the absolute median residual $|\tilde{\epsilon}|$ is $<0.01$~dex, corresponding to errors at the few-percent level. Also, the 16th--84th percentile scatter $\delta\epsilon$ is $\sim \pm 0.1$~dex, demonstrating the robustness of the predictions across our CNN configurations.

When we restrict the analysis to the relaxed cluster sample, the absolute median residuals increase slightly from $0.004$~dex (all sample) to $0.006$~dex (relaxed sample) for \cnnb, and from $0.006$~dex to $0.008$~dex for \cnnr. 
In contrast, the 16th--84th percentile scatter decreases modestly by $\sim 0.01$~dex. This behavior is also evident in Figure~\ref{fig:residual_distribution} as a small shift in the mean residual and a modest reduction in the standard deviation.

Adding the richness layer to the CNN has an effect on the mass predictions similar to restricting the analysis to the relaxed cluster sample. The absolute median residuals change only slightly, from $0.004$~dex in \cnnb\ to $0.006$~dex in \cnnr\ for the all sample, and from $0.006$~dex to $0.008$~dex for the relaxed sample. 
In contrast, the 16th--84th percentile scatter decreases substantially by $\sim 0.03$~dex, as also evident in Figure~\ref{fig:mvir_pred}. 
These results indicate that incorporating richness can significantly improve the mass inference and can outperform restricting the analysis to the relaxed sample, consistent with the strong correlation between richness and dynamical mass.

The standard deviation decreases from $\sigma_\epsilon = 0.133$  (all sample) to 0.124  (relaxed sample) for \cnnb and from 0.122 to 0.111 for \cnnr.  These reductions correspond to the tightening of the residual PDF in Figure~\ref{fig:residual_distribution}, where the $\pm 1\sigma$ spreads slightly shrink as the model and sample improves. Median residuals remain close to zero, indicating that differences are dominated by scatter and higher-order moments rather than bias.

Despite the reduced scatter, the residual distributions for the mass are strongly non-Gaussian. The skewness is consistently positive and large ($\gamma = 1.058$--$1.891$), indicating a pronounced right tail, while the excess kurtosis is very high ($\kappa = 4.773$--$10.74$), implying heavy tails. Overall, richness tends to tighten the scatter while tail-heaviness can remain substantial.

The architecture of \cnnb used in this study is based on
\citet{h2019}.  They reported that the prediction performance for all
cluster sample in the test range of $10^{14} \leq M_{\rm 200c} \leq
10^{15}$\hMsun is $\tilde\epsilon = -0.003$ and $\sigma_{\epsilon} =
0.132$~dex, where \mtzzc is the enclosed mass within a radius where
the spherical overdensity is 200 times the critical density.  Although
the halo mass definition differs, Table~\ref{table:pred_summary}
clearly shows that our \cnnb can predict the halo mass with comparable
performance of their model. Nevertheless, the mass range is broader in
our study than in \citet{h2019}, probably reflecting the quality of
training dataset used for the ML.  The Uchuu-UM catalog used in this
study is an eight-times larger volume and a five-times higher mass
resolution compared to the catalog used in \citet{h2019}.  
Indeed, when we restrict dataset with the same mass range
and optimize the network again, our CNN yields
$\tilde\epsilon=-0.002$ and $\sigma_{\epsilon}=0.114$~dex.  With the
same condition, our CNN can decrease $\sigma_{\epsilon}$ in the level
of 0.02~dex, reinforcing the importance of the quality of training
dataset.

\subsection{Inference of the scale radius}\label{sec:results:rs}
As indicated in Figure~\ref{fig:rs_pred} and
Table~\ref{table:pred_summary}, the absolute median residuals for
scale radius inference are comparable to those for mass, whereas the
scatter is larger for scale radius than for mass.  
In the case of the all sample, the median predicted scale radius in each true-value bin is unbiased and reproduces the true scale radius for \rs$\lesssim$ 300\hkpc. For \rs $\gtrsim$ 300\hkpc, the CNNs predict values that are slightly and systematically lower by $\sim$0.1~dex. This bias is not seen in the relaxed sample, which shows a $0.04$--$0.05$~dex reduction in the 16th--84th percentile scatter.

This bias can be understood as follows.  We adopt the scale radius
computed by numerically solving Equation \eqref{eq:vvir}. When a massive subhalo exists in the outskirts of a halo, as can occur immediately after a major merger, 
$V_{\rm max}$ tends to be close to $V_{\rm vir}$, and the resulting concentration is $c \simeq 2$ by definition.  
The minimum cluster mass considered in this study is $10^{13.5}$\hMsun,
corresponding to the scale radius of $\sim$300\hkpc.  Therefore, the
effect of these low-concentration halos appears primarily at
\rs$\gtrsim$ 300\hkpc, which is also visible in Figure~\ref{fig:cluster_func} as a slight enhancement in the number
density at this scale for the all sample.  This enhancement is absent
for the relaxed sample because halos with recent major mergers are
dynamically unrelaxed.  Consequently, the bias at \rs$\gtrsim$ 300\hkpc
is seen in the all sample but is removed in the relaxed sample.

Adding the richness layer in the CNN has a smaller impact than
restricting the analysis to the relaxed sample.  The absolute median
residuals are comparable across models, while the ranges of the
16th-84th percentile slightly decrease by the level of
0.01$\sim$0.02~dex. This reduction is smaller than the improvements obtained by selecting the relaxed sample by about a factor of three.  These results demonstrate that restricting to dynamically relaxed clusters yields a substantially larger improvement in scale radius inference than adding richness, in stark contrast to the mass inference.

Table~\ref{table:pred_summary} shows systematically larger scatter 
for the scale radius than for the mass, and the most
effective improvement comes from restricting the sample to dynamically
relaxed clusters. The standard deviation decreases from
$\sigma_\epsilon = 0.180$ in the all sample to 0.154 in the relaxed sample for \cnnb. Adding richness produces smaller additional improvements, which is consistent with the residual PDF in Figure~\ref{fig:residual_distribution} becoming noticeably narrower
when moving from the all sample to the relaxed sample. The distribution shape for \rs is comparatively well-behaved. Skewness stays near zero ($\gamma =-0.003$ to $0.230$) and excess kurtosis is modest ($\kappa = 0.759$ to $1.247$), suggesting that limitations in \rs inference are driven mainly by scatter rather than asymmetry or outliers.

\section{Conclusions and Discussions}\label{sec:sec5}
We have investigated CNN-based inference of the halo virial mass \mvir
and the scale radius \rs of galaxy clusters using mock observational
data constructed from the Uchuu--UniverseMachine catalog. In our
approach, each cluster is represented by a smoothed $48\times48$
phase-space density map of member galaxies 
that is designed to emulate
realistic membership selection, including contamination by
interlopers. We trained and tested two architectures: a baseline
network \cnnb following \citet{h2019} and an extended network \cnnr
that incorporates richness as an additional scalar input. We 
quantified their performance using the logarithmic residual
$\epsilon=\log(Y_{\rm pred}/Y_{\rm true})$, where $Y_{\rm {true}}$ and
$Y_{\rm {pred}}$ are the true and predicted values, respectively.

Over the test ranges $10^{13.7}\leq M_{\rm vir}\leq 10^{15.3}$\hMsun
and $10^{1.7}\leq r_{\rm s}\leq 10^{2.7}$\hkpc, all models achieve
nearly unbiased predictions with absolute median residuals within
0.01~dex. For the virial mass, adding richness clearly decreases the
1$\sigma$ scatter of $\epsilon$: it decreases from 0.133 to 0.122~dex
for the all sample, and from 0.124 to 0.111~dex for the relaxed sample. The
residual distribution for the virial mass, however, remains notably
non-Gaussian, with positive skewness and large excess kurtosis in all
configurations, indicating a pronounced high-residual tail even when
the 1$\sigma$ scatter is reduced.  For the scale radius, the inference
is intrinsically more challenging than for the virial mass, leading to
larger 1$\sigma$ scatter. In contrast to the mass inference, selecting
dynamically relaxed clusters is more effective than adding richness:
the 1$\sigma$ scatter improves from 0.180 to 0.154~dex for \cnnb, and
0.175 to 0.148~dex for \cnnr.

These results have an observational implication for how one should
select ``relaxed'' systems. In practice, the dynamical state of an
observed cluster is not directly known from projected phase-space
information of member galaxies alone, and observational samples can
include systems with ongoing mergers or substantial substructures. Our
findings suggest that such dynamically unstable systems can
systematically degrade the inference of the scale radius. 
Therefore, analyses focusing on halo structural parameters should either 
incorporate the impact of unrelaxed systems as an additional source of uncertainty
or restrict to clusters that are likely dynamically relaxed based on independent
observational indicators of dynamical state such as the prevalence of substructure, 
X-ray data, and the offset between the brightest cluster galaxy and the cluster center 
\citep[e.g.,][]{Dressler1988, Lopes2018, Yuan2022}. 
We will incorporate such indicators into the ML models and evaluate how 
the accuracy of concentration inference is improved in future work.

\begin{acknowledgments}
We thank Naohiro Manago for helpful discussions. 
This work has been supported by IAAR Research Support Program in Chiba
University Japan, MEXT/JSPS KAKENHI (Grant Number JP19KK0344 and
JP25H00662), MEXT as ``Program for Promoting Researches on the
Supercomputer Fugaku'' (JPMXP1020230406), and
JICFuS.  

We thank Instituto de Astrofisica de Andalucia (IAA-CSIC), Centro de
Supercomputacion de Galicia (CESGA) and the Spanish academic and
research network (RedIRIS) in Spain for hosting Uchuu DR1, DR2 and DR3
in the Skies \& Universes site for cosmological simulations. The Uchuu
simulations were carried out on Aterui II supercomputer at Center for
Computational Astrophysics, CfCA, of National Astronomical Observatory
of Japan, and the K computer at the RIKEN Advanced Institute for
Computational Science. The Uchuu Data Releases efforts have made use
of the skun@IAA\_RedIRIS and skun6@IAA computer facilities managed by
the IAA-CSIC in Spain (MICINN EU-Feder grant EQC2018-004366-P).
\end{acknowledgments}

\bibliographystyle{aasjournalv7}

\begin{thebibliography}{}
\expandafter\ifx\csname natexlab\endcsname\relax\def\natexlab#1{#1}\fi
\providecommand{\url}[1]{\href{#1}{#1}}
\providecommand{\dodoi}[1]{doi:~\href{http://doi.org/#1}{\nolinkurl{#1}}}
\providecommand{\doeprint}[1]{\href{http://ascl.net/#1}{\nolinkurl{http://ascl.net/#1}}}
\providecommand{\doarXiv}[1]{\href{https://arxiv.org/abs/#1}{\nolinkurl{https://arxiv.org/abs/#1}}}

\bibitem[{M.~H. {Abdullah} {et~al.}(2023){Abdullah}, {Wilson}, {Klypin}, \&
  {Ishiyama}}]{2023ApJ...955...26A}
{Abdullah}, M.~H., {Wilson}, G., {Klypin}, A., \& {Ishiyama}, T. 2023,
  \bibinfo{title}{{Constraining Cosmological Parameters Using the Cluster
  Mass-Richness Relation},} \apj, 955, 26, \dodoi{10.3847/1538-4357/ace773}

\bibitem[{S.~W. {Allen} {et~al.}(2011){Allen}, {Evrard}, \&
  {Mantz}}]{Allen2011}
{Allen}, S.~W., {Evrard}, A.~E., \& {Mantz}, A.~B. 2011,
  \bibinfo{title}{{Cosmological Parameters from Observations of Galaxy
  Clusters},} \araa, 49, 409, \dodoi{10.1146/annurev-astro-081710-102514}

\bibitem[{H. {Aung} {et~al.}(2023){Aung}, {Nagai}, {Klypin}, {Behroozi},
  {Abdullah}, {Ishiyama}, {Prada}, {P{\'e}rez}, {L{\'o}pez Cacheiro}, \&
  {Ruedas}}]{uchuu-um2023}
{Aung}, H., {Nagai}, D., {Klypin}, A., {et~al.} 2023, \bibinfo{title}{{The
  Uchuu-universe machine data set: galaxies in and around clusters},} \mnras,
  519, 1648, \dodoi{10.1093/mnras/stac3514}

\bibitem[{P. {Behroozi} {et~al.}(2019){Behroozi}, {Wechsler}, {Hearin}, \&
  {Conroy}}]{Behroozi2019}
{Behroozi}, P., {Wechsler}, R.~H., {Hearin}, A.~P., \& {Conroy}, C. 2019,
  \bibinfo{title}{{UNIVERSEMACHINE: The correlation between galaxy growth and
  dark matter halo assembly from z = 0-10},} \mnras, 488, 3143,
  \dodoi{10.1093/mnras/stz1182}

\bibitem[{P.~S. {Behroozi} {et~al.}(2013{\natexlab{a}}){Behroozi}, {Wechsler},
  \& {Wu}}]{Behroozi2013}
{Behroozi}, P.~S., {Wechsler}, R.~H., \& {Wu}, H.-Y. 2013{\natexlab{a}},
  \bibinfo{title}{{The ROCKSTAR Phase-space Temporal Halo Finder and the
  Velocity Offsets of Cluster Cores},} \apj, 762, 109,
  \dodoi{10.1088/0004-637X/762/2/109}

\bibitem[{P.~S. {Behroozi} {et~al.}(2013{\natexlab{b}}){Behroozi}, {Wechsler},
  {Wu}, {Busha}, {Klypin}, \& {Primack}}]{Behroozi2013b}
{Behroozi}, P.~S., {Wechsler}, R.~H., {Wu}, H.-Y., {et~al.} 2013{\natexlab{b}},
  \bibinfo{title}{{Gravitationally Consistent Halo Catalogs and Merger Trees
  for Precision Cosmology},} \apj, 763, 18, \dodoi{10.1088/0004-637X/763/1/18}

\bibitem[{A. {Biviano} {et~al.}(2006){Biviano}, {Murante}, {Borgani},
  {Diaferio}, {Dolag}, \& {Girardi}}]{Biviano2006}
{Biviano}, A., {Murante}, G., {Borgani}, S., {et~al.} 2006, \bibinfo{title}{{On
  the efficiency and reliability of cluster mass estimates based on member
  galaxies},} \aap, 456, 23, \dodoi{10.1051/0004-6361:20064918}

\bibitem[{G.~L. {Bryan} \& M.~L. {Norman}(1998){Bryan} \& {Norman}}]{Bryan1998}
{Bryan}, G.~L., \& {Norman}, M.~L. 1998, \bibinfo{title}{{Statistical
  Properties of X-Ray Clusters: Analytic and Numerical Comparisons},} \apj,
  495, 80, \dodoi{10.1086/305262}

\bibitem[{J.~S. {Bullock} {et~al.}(2001){Bullock}, {Kolatt}, {Sigad},
  {Somerville}, {Kravtsov}, {Klypin}, {Primack}, \& {Dekel}}]{Bullock2001}
{Bullock}, J.~S., {Kolatt}, T.~S., {Sigad}, Y., {et~al.} 2001,
  \bibinfo{title}{{Profiles of dark haloes: evolution, scatter and
  environment},} \mnras, 321, 559, \dodoi{10.1046/j.1365-8711.2001.04068.x}

\bibitem[{P. {Busch} \& S.~D.~M. {White}(2017){Busch} \& {White}}]{Busch2017}
{Busch}, P., \& {White}, S. D.~M. 2017, \bibinfo{title}{{Assembly bias and
  splashback in galaxy clusters},} \mnras, 470, 4767,
  \dodoi{10.1093/mnras/stx1584}

\bibitem[{B. {Diemer} \& M. {Joyce}(2019){Diemer} \& {Joyce}}]{Diemer2019}
{Diemer}, B., \& {Joyce}, M. 2019, \bibinfo{title}{{An Accurate Physical Model
  for Halo Concentrations},} \apj, 871, 168, \dodoi{10.3847/1538-4357/aafad6}

\bibitem[{A. {Dressler} \& S.~A. {Shectman}(1988){Dressler} \&
  {Shectman}}]{Dressler1988}
{Dressler}, A., \& {Shectman}, S.~A. 1988, \bibinfo{title}{{Evidence for
  Substructure in Rich Clusters of Galaxies from Radial-Velocity
  Measurements},} \aj, 95, 985, \dodoi{10.1086/114694}

\bibitem[{A.~E. {Evrard} {et~al.}(2008){Evrard}, {Bialek}, {Busha}, {White},
  {Habib}, {Heitmann}, {Warren}, {Rasia}, {Tormen}, {Moscardini}, {Power},
  {Jenkins}, {Gao}, {Frenk}, {Springel}, {White}, \& {Diemand}}]{Evrard2008}
{Evrard}, A.~E., {Bialek}, J., {Busha}, M., {et~al.} 2008,
  \bibinfo{title}{{Virial Scaling of Massive Dark Matter Halos: Why Clusters
  Prefer a High Normalization Cosmology},} \apj, 672, 122,
  \dodoi{10.1086/521616}

\bibitem[{L. {Gao} {et~al.}(2005){Gao}, {Springel}, \& {White}}]{Gao2005}
{Gao}, L., {Springel}, V., \& {White}, S. D.~M. 2005, \bibinfo{title}{{The age
  dependence of halo clustering},} \mnras, 363, L66,
  \dodoi{10.1111/j.1745-3933.2005.00084.x}

\bibitem[{L. {Gao} \& S.~D.~M. {White}(2007){Gao} \& {White}}]{Gao2007}
{Gao}, L., \& {White}, S. D.~M. 2007, \bibinfo{title}{{Assembly bias in the
  clustering of dark matter haloes},} \mnras, 377, L5,
  \dodoi{10.1111/j.1745-3933.2007.00292.x}

\bibitem[{A. Gonz\'alez(2010)Gonz\'alez}]{g2010}
Gonz\'alez, A. 2010, \bibinfo{title}{Measurement of Areas on a Sphere Using
  Fibonacci and Latitude?Longitude Lattices,} Math Geosci, 42, 49.
\newblock \url{https://doi.org/10.1007/s11004-009-9257-x}

\bibitem[{M. {Ho} {et~al.}(2021){Ho}, {Farahi}, {Rau}, \& {Trac}}]{Ho2021}
{Ho}, M., {Farahi}, A., {Rau}, M.~M., \& {Trac}, H. 2021,
  \bibinfo{title}{{Approximate Bayesian Uncertainties on Deep Learning
  Dynamical Mass Estimates of Galaxy Clusters},} \apj, 908, 204,
  \dodoi{10.3847/1538-4357/abd101}

\bibitem[{M. Ho {et~al.}(2019)Ho, Rau, Ntampaka, Farahi, Trac, \&
  Poczos}]{h2019}
Ho, M., Rau, M.~M., Ntampaka, M., {et~al.} 2019, \bibinfo{title}{A Robust and
  Efficient Deep Learning Method for Dynamical Mass Measurements of Galaxy
  Clusters,} The Astrophysical Journal, 887, 25,
  \dodoi{10.3847/1538-4357/ab4f82}

\bibitem[{H. {Hoekstra} {et~al.}(2013){Hoekstra}, {Bartelmann}, {Dahle},
  {Israel}, {Limousin}, \& {Meneghetti}}]{Hoekstra2013}
{Hoekstra}, H., {Bartelmann}, M., {Dahle}, H., {et~al.} 2013,
  \bibinfo{title}{{Masses of Galaxy Clusters from Gravitational Lensing},}
  \ssr, 177, 75, \dodoi{10.1007/s11214-013-9978-5}

\bibitem[{K. {Holhjem} {et~al.}(2009){Holhjem}, {Schirmer}, \&
  {Dahle}}]{Holhjem09}
{Holhjem}, K., {Schirmer}, M., \& {Dahle}, H. 2009, \bibinfo{title}{{Weak
  lensing density profiles and mass reconstructions of the galaxy clusters
  Abell 1351 and Abell 1995},} \aap, 504, 1, \dodoi{10.1051/0004-6361/20079006}

\bibitem[{T. {Ishiyama} {et~al.}(2009){Ishiyama}, {Fukushige}, \&
  {Makino}}]{Ishiyama2009}
{Ishiyama}, T., {Fukushige}, T., \& {Makino}, J. 2009,
  \bibinfo{title}{{Variation of the Subhalo Abundance in Dark Matter Halos},}
  \apj, 696, 2115, \dodoi{10.1088/0004-637X/696/2/2115}

\bibitem[{T. {Ishiyama} {et~al.}(2021){Ishiyama}, {Prada}, {Klypin}, {Sinha},
  {Metcalf}, {Jullo}, {Altieri}, {Cora}, {Croton}, {de la Torre},
  {Mill{\'a}n-Calero}, {Oogi}, {Ruedas}, \&
  {Vega-Mart{\'\i}nez}}]{Ishiyama2021}
{Ishiyama}, T., {Prada}, F., {Klypin}, A.~A., {et~al.} 2021,
  \bibinfo{title}{{The Uchuu simulations: Data Release 1 and dark matter halo
  concentrations},} \mnras, 506, 4210, \dodoi{10.1093/mnras/stab1755}

\bibitem[{A. {Jeeson-Daniel} {et~al.}(2011){Jeeson-Daniel}, {Dalla Vecchia},
  {Haas}, \& {Schaye}}]{Jeeson-Danie2011}
{Jeeson-Daniel}, A., {Dalla Vecchia}, C., {Haas}, M.~R., \& {Schaye}, J. 2011,
  \bibinfo{title}{{The correlation structure of dark matter halo properties},}
  \mnras, 415, L69, \dodoi{10.1111/j.1745-3933.2011.01081.x}

\bibitem[{N. {Kaiser}(1984){Kaiser}}]{Kaiser1984}
{Kaiser}, N. 1984, \bibinfo{title}{{On the spatial correlations of Abell
  clusters.},} \apjl, 284, L9, \dodoi{10.1086/184341}

\bibitem[{D.~P. Kingma \& J. Ba(2014)Kingma \& Ba}]{Kingma2014AdamAM}
Kingma, D.~P., \& Ba, J. 2014, \bibinfo{title}{Adam: A Method for Stochastic
  Optimization,} CoRR, abs/1412.6980.
\newblock \url{https://api.semanticscholar.org/CorpusID:6628106}

\bibitem[{A. {Klypin} {et~al.}(2016){Klypin}, {Yepes}, {Gottl{\"o}ber},
  {Prada}, \& {He{\ss}}}]{Klypin2016}
{Klypin}, A., {Yepes}, G., {Gottl{\"o}ber}, S., {Prada}, F., \& {He{\ss}}, S.
  2016, \bibinfo{title}{{MultiDark simulations: the story of dark matter halo
  concentrations and density profiles},} \mnras, 457, 4340,
  \dodoi{10.1093/mnras/stw248}

\bibitem[{A.~A. {Klypin} {et~al.}(2011){Klypin}, {Trujillo-Gomez}, \&
  {Primack}}]{Klypin2011}
{Klypin}, A.~A., {Trujillo-Gomez}, S., \& {Primack}, J. 2011,
  \bibinfo{title}{{Dark Matter Halos in the Standard Cosmological Model:
  Results from the Bolshoi Simulation},} \apj, 740, 102,
  \dodoi{10.1088/0004-637X/740/2/102}

\bibitem[{P.~A.~A. {Lopes} {et~al.}(2018){Lopes}, {Trevisan}, {Lagan{\'a}},
  {Durret}, {Ribeiro}, \& {Rembold}}]{Lopes2018}
{Lopes}, P. A.~A., {Trevisan}, M., {Lagan{\'a}}, T.~F., {et~al.} 2018,
  \bibinfo{title}{{Optical substructure and BCG offsets of Sunyaev-Zel'dovich
  and X-ray-selected galaxy clusters},} \mnras, 478, 5473,
  \dodoi{10.1093/mnras/sty1374}

\bibitem[{H. {Miyatake} {et~al.}(2016){Miyatake}, {More}, {Takada}, {Spergel},
  {Mandelbaum}, {Rykoff}, \& {Rozo}}]{Miyatake2016}
{Miyatake}, H., {More}, S., {Takada}, M., {et~al.} 2016,
  \bibinfo{title}{{Evidence of Halo Assembly Bias in Massive Clusters},} \prl,
  116, 041301, \dodoi{10.1103/PhysRevLett.116.041301}

\bibitem[{H.~J. {Mo} \& S.~D.~M. {White}(1996){Mo} \& {White}}]{Mo1996}
{Mo}, H.~J., \& {White}, S.~D.~M. 1996, \bibinfo{title}{{An analytic model for
  the spatial clustering of dark matter haloes},} \mnras, 282, 347,
  \dodoi{10.1093/mnras/282.2.347}

\bibitem[{S. {More} {et~al.}(2016){More}, {Miyatake}, {Takada}, {Diemer},
  {Kravtsov}, {Dalal}, {More}, {Murata}, {Mandelbaum}, {Rozo}, {Rykoff},
  {Oguri}, \& {Spergel}}]{Surhud2016}
{More}, S., {Miyatake}, H., {Takada}, M., {et~al.} 2016,
  \bibinfo{title}{{Detection of the Splashback Radius and Halo Assembly Bias of
  Massive Galaxy Clusters},} \apj, 825, 39, \dodoi{10.3847/0004-637X/825/1/39}

\bibitem[{V. Nair \& G.~E. Hinton(2010)Nair \& Hinton}]{ReLU}
Nair, V., \& Hinton, G.~E. 2010, \bibinfo{title}{Rectified linear units improve
  restricted boltzmann machines,} Madison, WI, USA: Omnipress

\bibitem[{J.~F. Navarro {et~al.}(1996)Navarro, Frenk, \& White}]{n1996}
Navarro, J.~F., Frenk, C.~S., \& White, S. D.~M. 1996, \bibinfo{title}{The
  Structure of Cold Dark Matter Halos,} The Astrophysical Journal, 462, 563,
  \dodoi{10.1086/177173}

\bibitem[{J.~F. Navarro {et~al.}(1997)Navarro, Frenk, \& White}]{n1997}
Navarro, J.~F., Frenk, C.~S., \& White, S. D.~M. 1997, \bibinfo{title}{A
  Universal Density Profile from Hierarchical Clustering,} The Astrophysical
  Journal, 490, 493, \dodoi{10.1086/304888}

\bibitem[{M. {Ntampaka} {et~al.}(2015){Ntampaka}, {Trac}, {Sutherland},
  {Battaglia}, {P{\'o}czos}, \& {Schneider}}]{Ntampaka2015}
{Ntampaka}, M., {Trac}, H., {Sutherland}, D.~J., {et~al.} 2015,
  \bibinfo{title}{{A Machine Learning Approach for Dynamical Mass Measurements
  of Galaxy Clusters},} \apj, 803, 50, \dodoi{10.1088/0004-637X/803/2/50}

\bibitem[{M. {Ntampaka} {et~al.}(2016){Ntampaka}, {Trac}, {Sutherland},
  {Fromenteau}, {P{\'o}czos}, \& {Schneider}}]{Ntampaka2016}
{Ntampaka}, M., {Trac}, H., {Sutherland}, D.~J., {et~al.} 2016,
  \bibinfo{title}{{Dynamical Mass Measurements of Contaminated Galaxy Clusters
  Using Machine Learning},} \apj, 831, 135, \dodoi{10.3847/0004-637X/831/2/135}

\bibitem[{M. {Oguri} {et~al.}(2012){Oguri}, {Bayliss}, {Dahle}, {Sharon},
  {Gladders}, {Natarajan}, {Hennawi}, \& {Koester}}]{Oguri2012}
{Oguri}, M., {Bayliss}, M.~B., {Dahle}, H., {et~al.} 2012,
  \bibinfo{title}{{Combined strong and weak lensing analysis of 28 clusters
  from the Sloan Giant Arcs Survey},} \mnras, 420, 3213,
  \dodoi{10.1111/j.1365-2966.2011.20248.x}

\bibitem[{ {Planck Collaboration} {et~al.}(2016){Planck Collaboration}, {Ade},
  {Aghanim}, {Arnaud}, {Ashdown}, {Aumont}, {Baccigalupi}, {Banday},
  {Barreiro}, {Bartlett}, \& et~al.}]{Planck2016}
{Planck Collaboration}, {Ade}, P.~A.~R., {Aghanim}, N., {et~al.} 2016,
  \bibinfo{title}{{Planck 2015 results. XIII. Cosmological parameters},} \aap,
  594, A13, \dodoi{10.1051/0004-6361/201525830}

\bibitem[{F. {Prada} {et~al.}(2012){Prada}, {Klypin}, {Cuesta},
  {Betancort-Rijo}, \& {Primack}}]{Prada2012}
{Prada}, F., {Klypin}, A.~A., {Cuesta}, A.~J., {Betancort-Rijo}, J.~E., \&
  {Primack}, J. 2012, \bibinfo{title}{{Halo concentrations in the standard
  {$\Lambda$} cold dark matter cosmology},} \mnras, 423, 3018,
  \dodoi{10.1111/j.1365-2966.2012.21007.x}

\bibitem[{G.~W. {Pratt} {et~al.}(2019){Pratt}, {Arnaud}, {Biviano}, {Eckert},
  {Ettori}, {Nagai}, {Okabe}, \& {Reiprich}}]{Pratt2019}
{Pratt}, G.~W., {Arnaud}, M., {Biviano}, A., {et~al.} 2019,
  \bibinfo{title}{{The Galaxy Cluster Mass Scale and Its Impact on Cosmological
  Constraints from the Cluster Population},} \ssr, 215, 25,
  \dodoi{10.1007/s11214-019-0591-0}

\bibitem[{G.~W. {Pratt} {et~al.}(2009){Pratt}, {Croston}, {Arnaud}, \&
  {B{\"o}hringer}}]{Pratt2009}
{Pratt}, G.~W., {Croston}, J.~H., {Arnaud}, M., \& {B{\"o}hringer}, H. 2009,
  \bibinfo{title}{{Galaxy cluster X-ray luminosity scaling relations from a
  representative local sample (REXCESS)},} \aap, 498, 361,
  \dodoi{10.1051/0004-6361/200810994}

\bibitem[{E. {Rozo} {et~al.}(2009){Rozo}, {Rykoff}, {Evrard}, {Becker},
  {McKay}, {Wechsler}, {Koester}, {Hao}, {Hansen}, {Sheldon}, {Johnston},
  {Annis}, \& {Frieman}}]{Rozo2009}
{Rozo}, E., {Rykoff}, E.~S., {Evrard}, A., {et~al.} 2009,
  \bibinfo{title}{{Constraining the Scatter in the Mass-richness Relation of
  maxBCG Clusters with Weak Lensing and X-ray Data},} \apj, 699, 768,
  \dodoi{10.1088/0004-637X/699/1/768}

\bibitem[{E.~S. {Rykoff} {et~al.}(2012){Rykoff}, {Koester}, {Rozo}, {Annis},
  {Evrard}, {Hansen}, {Hao}, {Johnston}, {McKay}, \& {Wechsler}}]{Rykoff2012}
{Rykoff}, E.~S., {Koester}, B.~P., {Rozo}, E., {et~al.} 2012,
  \bibinfo{title}{{Robust Optical Richness Estimation with Reduced Scatter},}
  \apj, 746, 178, \dodoi{10.1088/0004-637X/746/2/178}

\bibitem[{E.~S. {Rykoff} {et~al.}(2014){Rykoff}, {Rozo}, {Busha}, {Cunha},
  {Finoguenov}, {Evrard}, {Hao}, {Koester}, {Leauthaud}, {Nord}, {Pierre},
  {Reddick}, {Sadibekova}, {Sheldon}, \& {Wechsler}}]{Rykoff2014}
{Rykoff}, E.~S., {Rozo}, E., {Busha}, M.~T., {et~al.} 2014,
  \bibinfo{title}{{redMaPPer. I. Algorithm and SDSS DR8 Catalog},} \apj, 785,
  104, \dodoi{10.1088/0004-637X/785/2/104}

\bibitem[{D. Scott(2015)Scott}]{sc2015}
Scott, D. 2015, \bibinfo{title}{Multivariate Density Estimation: Theory,
  Practice, and Visualization,} Wiley.
\newblock \url{https://books.google.co.jp/books?id=XZ03BwAAQBAJ}

\bibitem[{M. {Simet} {et~al.}(2017){Simet}, {McClintock}, {Mandelbaum}, {Rozo},
  {Rykoff}, {Sheldon}, \& {Wechsler}}]{Simet2017}
{Simet}, M., {McClintock}, T., {Mandelbaum}, R., {et~al.} 2017,
  \bibinfo{title}{{Weak lensing measurement of the mass-richness relation of
  SDSS redMaPPer clusters},} \mnras, 466, 3103, \dodoi{10.1093/mnras/stw3250}

\bibitem[{R.~A. {Skibba} \& A.~V. {Macci{\`o}}(2011){Skibba} \&
  {Macci{\`o}}}]{Skibba2011}
{Skibba}, R.~A., \& {Macci{\`o}}, A.~V. 2011, \bibinfo{title}{{Properties of
  dark matter haloes and their correlations: the lesson from principal
  component analysis},} \mnras, 416, 2388,
  \dodoi{10.1111/j.1365-2966.2011.19218.x}

\bibitem[{J. {Soltis} {et~al.}(2025){Soltis}, {Ntampaka}, {Diemer}, {ZuHone},
  {Bose}, {Delgado}, {Hadzhiyska}, {Hern{\'a}ndez-Aguayo}, {Nagai}, \&
  {Trac}}]{Soltis2025}
{Soltis}, J., {Ntampaka}, M., {Diemer}, B., {et~al.} 2025, \bibinfo{title}{{A
  Multiwavelength Technique for Estimating Galaxy Cluster Mass Accretion
  Rates},} \apj, 985, 212, \dodoi{10.3847/1538-4357/adcfa4}

\bibitem[{T. {Sunayama} \& S. {More}(2019){Sunayama} \& {More}}]{Sunayama2019}
{Sunayama}, T., \& {More}, S. 2019, \bibinfo{title}{{On the measurements of
  assembly bias and splashback radius using optically selected galaxy
  clusters},} \mnras, 490, 4945, \dodoi{10.1093/mnras/stz2832}

\bibitem[{T. {Tokuue} {et~al.}(2024){Tokuue}, {Ishiyama}, {Osato}, {Tanaka}, \&
  {Behroozi}}]{Tokuue2024}
{Tokuue}, T., {Ishiyama}, T., {Osato}, K., {Tanaka}, S., \& {Behroozi}, P.
  2024, \bibinfo{title}{{MPI-Rockstar: a Hybrid MPI and OpenMP Parallel
  Implementation of the Rockstar Halo finder},} arXiv e-prints,
  arXiv:2412.18629, \dodoi{10.48550/arXiv.2412.18629}

\bibitem[{A. {Vikhlinin} {et~al.}(2009){Vikhlinin}, {Burenin}, {Ebeling},
  {Forman}, {Hornstrup}, {Jones}, {Kravtsov}, {Murray}, {Nagai}, {Quintana}, \&
  {Voevodkin}}]{Vikhlinin2009}
{Vikhlinin}, A., {Burenin}, R.~A., {Ebeling}, H., {et~al.} 2009,
  \bibinfo{title}{{Chandra Cluster Cosmology Project. II. Samples and X-Ray
  Data Reduction},} \apj, 692, 1033, \dodoi{10.1088/0004-637X/692/2/1033}

\bibitem[{L. {Wang} \& P.~J. {Steinhardt}(1998){Wang} \&
  {Steinhardt}}]{Wang1998}
{Wang}, L., \& {Steinhardt}, P.~J. 1998, \bibinfo{title}{{Cluster Abundance
  Constraints for Cosmological Models with a Time-varying, Spatially
  Inhomogeneous Energy Component with Negative Pressure},} \apj, 508, 483,
  \dodoi{10.1086/306436}

\bibitem[{R.~H. {Wechsler} {et~al.}(2002){Wechsler}, {Bullock}, {Primack},
  {Kravtsov}, \& {Dekel}}]{Wechsler2002}
{Wechsler}, R.~H., {Bullock}, J.~S., {Primack}, J.~R., {Kravtsov}, A.~V., \&
  {Dekel}, A. 2002, \bibinfo{title}{{Concentrations of Dark Halos from Their
  Assembly Histories},} \apj, 568, 52, \dodoi{10.1086/338765}

\bibitem[{R.~H. {Wechsler} {et~al.}(2006){Wechsler}, {Zentner}, {Bullock},
  {Kravtsov}, \& {Allgood}}]{Wechsler2006}
{Wechsler}, R.~H., {Zentner}, A.~R., {Bullock}, J.~S., {Kravtsov}, A.~V., \&
  {Allgood}, B. 2006, \bibinfo{title}{{The Dependence of Halo Clustering on
  Halo Formation History, Concentration, and Occupation},} \apj, 652, 71,
  \dodoi{10.1086/507120}

\bibitem[{G. {Wilson} {et~al.}(1996){Wilson}, {Cole}, \& {Frenk}}]{Wilson96}
{Wilson}, G., {Cole}, S., \& {Frenk}, C.~S. 1996, \bibinfo{title}{{Cluster mass
  reconstruction from weak gravitational lensing},} \mnras, 280, 199,
  \dodoi{10.1093/mnras/280.1.199}

\bibitem[{Z.~S. {Yuan} {et~al.}(2022){Yuan}, {Han}, \& {Wen}}]{Yuan2022}
{Yuan}, Z.~S., {Han}, J.~L., \& {Wen}, Z.~L. 2022, \bibinfo{title}{{Dynamical
  state of galaxy clusters evaluated from X-ray images},} \mnras, 513, 3013,
  \dodoi{10.1093/mnras/stac1037}

\bibitem[{Y. {Zu} {et~al.}(2017){Zu}, {Mandelbaum}, {Simet}, {Rozo}, \&
  {Rykoff}}]{Zu2017}
{Zu}, Y., {Mandelbaum}, R., {Simet}, M., {Rozo}, E., \& {Rykoff}, E.~S. 2017,
  \bibinfo{title}{{On the level of cluster assembly bias in SDSS},} \mnras,
  470, 551, \dodoi{10.1093/mnras/stx1264}

\end{thebibliography}

\end{document}